\documentclass[nomathfonts]{aipproc}
\layoutstyle{6x9}

\begin{document}

\title[Gamma-Ray Burst Intensities]{Intensity Distributions of Gamma-Ray Bursts}

\author{David L. Band}{address={X-2, Los Alamos National Laboratory,
Los Alamos, NM  87545 USA}}

\begin{abstract}
Observations of individual bursts chosen by the vagaries of
telescope availability demonstrated that bursts are \emph{not}
standard candles and that their apparent energy can be as great as
$10^{54}$~erg.  However, determining the distribution of their
apparent energy (and of other burst properties) requires the
statistical analysis of a well-defined burst sample; the sample
definition includes the threshold for including a burst in the
sample. Thus optical groups need to the criteria behind the
decision to search for a spectroscopic redshift. Currently the
burst samples are insufficient to choose between lognormal and
power law functional forms of the distribution, and the parameter
values for these functional forms differ between burst samples.
Similarly, the actual intensity distribution may be broader than
observed, with a low energy tail extending below the detection
threshold.
\end{abstract}

\maketitle

\section{Introduction}
Major advances in the study of gamma-ray bursts have resulted both
from the construction and analysis of carefully selected burst
samples, and from fortuitous discoveries of key properties of
individual bursts.  The BATSE burst sample, with precisely defined
detection thresholds, showed that the peak flux distribution and
the isotropic angular distribution were inconsistent with an
origin in the Galactic plane\cite{mee92} and marginally
inconsistent with a Galactic halo progenitor population, and thus
before 1997 most of the burst community concluded by a process of
elimination that bursts occur at cosmological distances. On the
other hand, beginning in 1997 the arcsecond localization of a few
bursts coincident with faint galaxies became the definitive
confirmation that bursts are indeed a cosmological phenomenon;
these bursts were chosen largely on the basis of the available
telescopes and the willingness of observers with access to these
telescopes to devote their precious observing time to following
these bursts. (Note that the possibility still remains that some
bursts---such as bursts with durations under $\sim1$~s---originate
from less distant sources.) Thus some of the major advances in the
study of bursts have resulted from the statistical analysis of
well-constructed burst samples, while others (often the more
spectacular breakthroughs) have followed observations of
individual bursts selected through the vagaries of the available
telescopes.  Here I discuss the interplay of statistical analysis
and serendipitous discovery in the determination of bursts'
intrinsic intensity.  I argue that observations of individual
bursts established the qualitative characteristics of burst
intensities but that determining the quantitative properties
requires well-defined criteria for choosing the bursts which
should be studied.  In particular, a multi-wavelength effort
involving different telescopes is required for determining
properties such as a burst's intensity and redshift:  an X-ray
burst detector (e.g., {\it Beppo-Sax}'s Wide Field Camera) first
localizes the burst to $\sim 3$~arcminute; an X-ray telescope
(e.g., {\it Beppo-Sax}'s Narrow Field Instrument) then localizes
the afterglow to about an arcminute; an imaging optical telescope
localizes the optical afterglow to less than an arcsecond; and
finally, optical spectroscopy determines the burst redshift
through the detection of either absorption lines in the afterglow
spectrum or emission lines from the host galaxy. The most
restrictive intensity threshold for mounting any component of this
campaign must be defined and reported to construct a well-defined
statistical sample.

\section{Constant Candle Bursts}
After the statistical analysis of the BATSE peak flux and spatial
distributions favored a cosmological origin, the burst community
by and large adopted the ``minimal cosmological
model''\cite{bah98} wherein bursts were standard candles and the
shape of the intensity distribution (e.g., for the peak photon
flux) resulted from the cosmological curvature of space. Although
this paradigm was expected to be overly simplistic, it was the
basis for Schaefer's conclusion\cite{sch92} that the predicted
host galaxies were missing from the best burst localizations then
available. Schaefer's assertion was contested\cite{lar96}, but the
statistical analysis of Band and Hartmann\cite{bah98} verified
that the \emph{expected} host galaxies were indeed absent.  Band
and Hartmann considered the galaxy detection thresholds in each
burst error box as well as the presence of the host and unrelated
background galaxies. However, testing the cosmological paradigm
which motivated Schaefer's search for the host galaxies became
moot with the observed coincidence of \emph{faint} galaxies with
afterglows localized to an arcsecond. As Schaefer had reported,
and Band and Hartmann corroborated, the minimal cosmological model
was indeed incorrect, not because bursts are not cosmological, but
because the model was too ``minimal:'' the deviation of the
cumulative intensity distribution from the $-3/2$ power law
expected for a homogeneous population in three-dimensional
Euclidean space does not result solely from the cosmological
curvature of space, and bursts are farther than derived by the
minimal model. Band, Hartmann and Schaefer\cite{bhs98} used the
statistical methodology of Band and Hartmann to find the burst
energy for which the host galaxy detections (or upper limits) were
consistent with the expected galaxy luminosity function. They
found a standard candle total burst energy of $\sim10^{53}$~erg.
The observed redshifts and energy fluences did give some burst
energies of this magnitude; however, because a range of burst
energies spanning $\sim 3$ decades was observed, these
observations also demonstrated that bursts are \emph{not} standard
candles in terms of the apparent total energy.

Thus the statistical analyses of Band and Hartmann\cite{bah98} and
Band, Hartmann and Schaefer\cite{bhs98} were appropriate for
answering questions such as whether the minimal cosmological model
was valid and what is the standard candle energy.  However, these
questions are answered more directly by a few observations of
randomly selected bursts:  only two bursts with significantly
different energies demonstrate that bursts are not standard
candles; only one bright burst with a faint host galaxy (or with
$z>1$) shows that the intensity distribution cannot be explained
solely by cosmological curvature; and a few burst energies give
the typical energy scale.

\section{The Intensity Distribution}
What is the distribution of the total energy emitted by a
gamma-ray burst?  We only observe the energy emitted in our
direction, which can be expressed as the \emph{apparent} total
energy, the energy emitted if the burst actually radiated
gamma-rays isotropically.  The apparent total energy is the actual
energy divided by the beaming fraction, the fraction of the sky
into which the gamma-rays are actually emitted; both the total
energy and the beaming fraction are relevant for understanding
burst physics. Indeed, Frail {\it et al.}\cite{fra01} calculated
the beaming fraction for a burst sample based on afterglow
evolution, and found a narrow distribution of the total energy
centered around $10^{51}$~erg.

Different measures of burst intensity have been used, such as the
peak energy luminosity, the total gamma-ray energy, the peak
photon flux, the total number of photons emitted, or the total
afterglow energy.  Note that bursts can be standard candles for at
most one of these intensity measures.  The intensity measure
studied is a matter of theoretical prejudice and ease of
calculation (the data may not be available to calculate some
measures).  Many studies have used the peak photon flux because
burst detectors, such as BATSE, trigger on the peak count rate,
and consequently the detection threshold for the peak flux is
fairly sharp.  However, I prefer the total energy emitted; the
observable is a burst's energy fluence.  In the current
theoretical scenario (see Piran\cite{pir00} for a review of
current burst theories) the gamma-ray emission results from
internal shocks when regions in a relativistic outflow with
different Lorentz factors collide.  The total gamma-ray energy
emitted should be related to the energy of the outflow while the
peak luminosity is a consequence of a particular internal shock,
which will result from the burst-specific distribution of Lorentz
factors within the outflow.  Thus I suspect that the emitted total
energy is fairly representative of the energy released while the
peak luminosity is more contingent on the details of the energy
release.

In studying the burst intensity distribution, my collaborators and
I have assumed specific functional forms. Thus Jimenez, Band and
Piran\cite{jim01} fit lognormal distributions to the total
apparent energy, the peak gamma-ray luminosity, and the total
X-ray afterglow energy, while recently I fit both lognormal and
simple power law distributions to the total apparent
energy.\cite{ban01}  When normalized to unity, these distributions
are the probability $p(I)$ that a burst has a given intensity $I$.
However, the probability $p_{\rm obs}(I)$ of \emph{observing} a
burst with a given intensity is that part of the intensity
distribution above the threshold for including the burst in our
sample.  For example, a high redshift burst will be detected, and
thus included in our sample, only if it was drawn from the high
end of the intensity distribution.  In both studies likelihood
functions were constructed from the probabilities $p_{\rm obs}(I)$
of obtaining each member of the sample.  The parameters of the
distribution function were determined by maximizing the
likelihood, and the parameters' confidence ranges were determined
by integrating over the likelihood surface.

These studies considered different burst samples.  Both Jimenez
{\it et al.}\cite{jim01} and my recent study\cite{ban01} used a
sample of 9 bursts with spectroscopic redshifts and BATSE spectra.
The energy fluences were calculated by fitting the GRB spectral
function\cite{ban93} to the BATSE spectra, and then integrating
the fits over the 20--2000~keV energy band and the burst duration.
In my study\cite{ban01} I also considered the 17 burst sample of
Frail {\it et al.}\cite{fra01} which adds bursts observed by {\it
Beppo-SAX}, {\it Ulysses}, KONUS and {\it NEAR} to the 9 BATSE
bursts.  For these two samples the detection thresholds are
unknown since the criteria for attempting to localize the bursts
and determine their redshifts have not been reported.  Finally, in
my study I used the 220 BATSE bursts with redshifts determined
through the conjectured correlation between light curve
variability and peak burst luminosity\cite{fen01}; this sample has
a well-defined threshold for including a burst.

Jimenez {\it et al.}\cite{jim01} extended their sample of bursts
with spectroscopic redshifts by including bursts with only a host
galaxy magnitude.  A redshift probability distribution can be
derived from a host galaxy magnitude using an empirical galaxy
redshift distribution and assuming a model for the rate at which
bursts occur in galaxies. Galaxy surveys such as the Hubble Deep
Field weight each detected galaxy equally, yet in most burst
scenarios bursts occur preferentially in massive or luminous
galaxies. Jimenez {\it et al.} tested various weighting schemes
using a sample of 10 bursts with both host galaxy magnitudes and
spectroscopic redshifts.  The test consisted of calculating a
likelihood using the redshift probability distributions evaluated
at the observed spectroscopic redshifts. The redshift probability
distribution used in this test should be modified to include only
the redshift range over which the observations could have
determined the redshift:  one of the few spectral lines detectable
in the spectra of faint galaxies must have been redshifted into
the telescope's bandpass. Once again a detection threshold is
required for the statistical analysis of a burst sample.  We found
that weighting the empirical galaxy redshift distribution (derived
from the Hubble Deep Field) by the host galaxy's luminosity at the
time of the burst was favored. This result is relevant to
progenitor scenarios (e.g., the galaxy luminosity for $z>1$ may be
proportional to the star formation rate, consistent with the
progenitors being massive, short-lived stars), although a proper
analysis requires sophisticated modeling of a galaxy's luminosity
history.

In my study of the burst energy\cite{ban01} I found that the
lognormal and power law distribution functions are both acceptable
descriptions of the data because the average of the cumulative
probability is consistent with the expected value of 1/2 within
the uncertainty resulting from the sample size.  The Bayesian odds
ratio demonstrated that neither function was favored over the
other (odds ratio of $\sim1$) for the two small burst samples with
spectroscopic redshifts, but the lognormal function is favored
(odds ratio of $\sim10^4$) for the large sample with redshifts
derived from the variability-luminosity correlation. The three
samples give significantly different best-fit parameter values,
which may result from the small sample sizes, the poorly
determined detection thresholds for the spectroscopic redshift
samples, and the uncertain validity of the variability-luminosity
correlation. For example, for the central energy of the lognormal
distribution I find: $E_0=1.3\times10^{53}$~erg (with a 90\%
confidence range of 0.016--3.2$\times 10^{53}$~erg) for the 9
burst BATSE sample; $E_0=5.2\times10^{53}$~erg (0.016--1.0$\times
10^{53}$~erg 90\% confidence range) for the 17 burst Frail {\it et
al.}\cite{fra01} sample; and $E_0=0.12\times10^{53}$~erg
(0.02--0.23$\times 10^{53}$~erg 90\% confidence range) for the 220
burst variability-luminosity\cite{fen01} sample.  For all three
samples the logarithmic width is nearly the same (the standard
deviation of $\ln E$ is $\sim2$).

The likelihood contours for the 2 parameters of the lognormal
distribution have a ridge of high likelihood running from the peak
towards lower central energy $E_0$ and larger logarithmic width
$\sigma$.  Similarly, the low energy cutoff for a simple power law
cannot be determined---we only know that it is below the lowest
energy observed.  Thus for both distribution functions the
likelihood does not rule out the possibility that the actual
energy distribution is wider than observed, with the low end
unobservable because of the detection threshold.  Only a larger
burst sample sensitive to fainter bursts will determine the extent
of the actual energy distribution function.

Implicit in this statistical methodology are assumptions about the
burst sample.  I assume that the energy distribution does not
evolve with redshift; this can eventually be tested by subdividing
a larger sample.  Further, I assume that there is no correlation
between a burst's energy and the ability to determine its
redshift; a burst's intrinsic intensity is assumed to be unrelated
to environmental factors which promote or suppress the afterglow
necessary to localize the burst.

As mentioned above, an accurate determination of the energy
distribution requires the fluence threshold for including a burst
in the sample.  In general, the statistical determination of a
burst property's distribution requires an understanding of the
thresholds for all observational steps.  In the case of the energy
distribution, the most restrictive threshold is the determination
of the redshift.  A decision was made that: a) the burst was
bright enough to attempt an afterglow detection; b) the afterglow
was sufficiently well-localized to find a host galaxy; and c) the
afterglow or the host galaxy are bright enough for spectroscopic
observations.  Thus, clear criteria are needed for the systematic
ground-based observations which will follow-up the large number of
well-localized bursts anticipated from {\it HETE-II} and {\it
SWIFT}.  Only with an understanding of the detection thresholds
will burst property distributions be determined quantitatively.
\begin{theacknowledgments}
I thank Dieter Hartmann for his comments on this paper.  This work
was performed under the auspices of the U.S. Department of Energy
by the Los Alamos National Laboratory under Contract No. W-7405-Eng-36.
\end{theacknowledgments}

\end{document}